\documentclass[runningheads]{svmult}

\usepackage{makeidx}   
\usepackage{graphicx}  
\usepackage{subeqnar}  
\usepackage{multicol}  
\usepackage{physprbb}  
\makeindex             

\begin{document}
\title*{The FALCON concept: multi-object spectroscopy combined with MCAO in near-IR}
%
%
%
%
\titlerunning{The FALCON concept}
%
\author{Fran\c{c}ois Hammer\inst{1}
\and Fr\'ed\'eric Say\`ede\inst{1}
\and Eric Gendron\inst{1}
\and Thierry Fusco\inst{2}
\and Denis Burgarella\inst{3}
\and V\'eronique Cayatte\inst{1}
\and Jean-Marc Conan\inst{2}
\and Fr\'ed\'eric Courbin \inst{1,4}
\and Hector Flores\inst{1}
\and Isabelle Guinouard\inst{1}
\and Laurent Jocou\inst{1}
\and Ariane Lan\c{c}on\inst{6} 
\and Guy Monnet\inst{5}
\and Mustapha Mouhcine\inst{6}
\and Fran\c{c}ois Rigaud\inst{1}
\and Daniel Rouan\inst{1}
\and G\'erard Rousset\inst{2}
\and V\'eronique Buat\inst{3}
\and Fr\'ed\'eric Zamkotsian\inst{3}}
\authorrunning{Fran\c{c}ois Hammer et al.}
%
%
\institute{Observatoire de Paris
\and ONERA
\and Laboratoire d'Astrophysique de Marseille
\and PUC, Depto. Astronomia y Astrofisica, Av. Vicu\~na Mackenna 4860, Santiago, Chile
\and ESO-Garching
\and Observatoire de Strasbourg}

\maketitle              

\begin{abstract}
\index{abstract} A large fraction  of the present-day stellar mass was
formed  between $z=0.5$  and $z\sim  3$ and  our understanding  of the
formation mechanisms at  work at these epochs requires  both high {\it
spatial}  and   high  {\it   spectral}  resolution:  one   shall  {\it
simultaneously} obtain  images of objects with typical  sizes as small
as 1-2  kpc ($\sim$ 0''.1),  while achieving 20-50 km/s  (R$\ge$ 5000)
spectral resolution. In addition,  the redshift range to be considered
implies that  most important spectral  features are redshifted  in the
near-infrared.  The obvious instrumental solution to adopt in order to
tackle the science goal is  therefore a combination of multi-object 3D
spectrograph with multi-conjugate adaptive  optics in large fields.  A
very promising way  to achieve such a technically  challenging goal is
to  relax  the conditions  of  the  traditional  full adaptive  optics
correction.   A partial,  but  still competitive  correction shall  be
prefered,  over a  much  wider field  of  view. This  can  be done  by
estimating the turbulent  volume from sets of natural  guide stars, by
optimizing the correction  to several and discrete small  areas of few
$arcsec^{2}$ selected  in a large  field (Nasmyth field of  25 arcmin)
and  by correcting  up to  the  6th, and eventually, up to the
60$^{th}$  Zernike modes.  Simulations on real extragalactic fields,
show that for most sources  ($>$ 80\%), the recovered resolution could
reach 0".15-0".25 in  the $J$ and $H$ bands. Detection of point-like
objects is improved  by factors from 3 to  $\ge$10, when compared with
an  instrument without adaptive  correction. The  proposed instrument
concept, FALCON, is equiped  with deployable mini-integral field units
(IFUs), achieving  spectral resolutions between R=5000  and 20000. Its
multiplex  capability,   combined  with  high spatial  and  spectral
resolution characteristics,  is a  natural ground based  complement to
the next generation of space telescopes. Galaxy formation in the early
Universe is  certainly a  main science driver.   We describe  here how
FALCON shall allow to answer puzzling questions in this area, although
the science cases naturally accessible to the instrument concept makes
it of interest for most areas of astrophysics.
\end{abstract}

\section{Scientific drivers for FALCON}

The second  generation of  instruments for the  VLT will  certainly be
driven  by  recurrent  and   still  unresolved  questions,  for  which
observational answers  are currently at the edge  of the possibilities
of 8 meter class telescopes.  The coming decade, shall be dedicated to
the understanding  of physical processes involved in  known targets from
well defined samples of targets. This will require observations of the
highest image quality and depth for  a limited number of sources.

 
\subsection{How and when galaxies formed ?}
The deepest past and current galaxy surveys (CFRS, HDF, DEEP) all show
that  galaxies beyond  $z=0.5$ were  smaller, more  irregular  and had
higher  star formation rate  than present-day  galaxies (Lilly  et al,
1998; Hammer  et al, 2001). Merging  rates were also  much higher than
today, being proportional to  $(1+z)^{4}$ (Yee \& Ellingston, 1995; Le
F\`evre  et  al, 2000).   Half  of  the  total light  contributing  to
infrared background has  been resolved by ISO (Elbaz  et al, 1999). It
was found to  be dominated by large disks  at $z=0.5-1.2$, often found
in interacting systems (Flores et  al, 1999). Accounting for all their
UV to IR and radio emission lead to a universal star formation density
which declines  by a factor  between 5 and  10 since $z=1$.   A simple
integration of  the global star  formation history shows that  half of
the present-day stars have been  formed since $z=1-1.5$ (Flores et al,
1999; Madau and Pozzetti, 2000; Madau et al, 2001, Franceschini et al,
2001). At  $z=3$, less than a  third of the  present-day metal content
was formed, even accounting  for large dust corrections (scenarii with
a constant  star formation density beyond $z\sim1$).   This is somehow
in contradiction with the  primordial collapse scenario, where massive
galaxies form at much earlier  epochs ($z\ge3$ or higher).  The latter
scenario  is supported  by the  apparent non-evolution  of  the number
density of giant  ellipticals at $z=1$ or higher  (Schade et al, 1999;
Cimatti, 2001,  this proceedings) and  by the small dispersion  of the
fundamental plane  for nearby ellipticals.   The debate is  still open
and  improved VLT  instrumentation  shall be  developed  to study  the
importance of  the evolutionary phase at $z=1-1.5$,  in a quantitative
and systematic way.\\
 
The  way galaxies  are assembling,  and how  they  are re-distributing
their mass,  velocities and angular momentum, is  largely unknown 
(e.g. Combes,  1999).   Cinematics and  chemistry of  galaxies
should be studied  over a variety of different  redshifts in order (1)
to determine  how common/important the  merging phenomenon is,  (2) to
map the distribution of disks  and spheroids at various times, and (3)
to firmly establish  the origin of the Hubble  sequence.  An important
redshift  range  for studying  galaxy  formation  is situated  between
$z=0.5$ and $z=2-3$, covering epochs for which most of the present-day
stars have  been formed.  At  these redshifts, the  important spectral
features such as  emission lines from [OII]3727 to  $H\alpha$, and the
stellar absorption lines, are observed  in the wavelength range $0.7 <
\lambda < 1.9$ microns where  ground-based 8-m telescopes are not only
competitive   with   future   space   telescopes   (NGST)   but   also
complementary.  Because of the  complexity of the mechanisms involved,
galaxy   formation  studies   require   simultaneously  high   spatial
resolution (down  to $\sim$  1-2kpc at $z>1$) and  moderate spectral
resolution (down to  $\sim$ 30 km/s). In other  words, 3D spectroscopy
is required, with good spatial sampling of the velocity field. This is
achievable   by  combining  3D   spectroscopy  with   adaptive  optics
correction over a large field of view.\\

At higher redshift, very little is known on the formation of the first
gravitationally   bound   structures,   namely   the   first   massive
proto-galaxies.  The  answer to the question is  intimately related to
the universal metal  formation history , as most  of the present-day
metal content is  locked into bulges of massive  galaxies (Fukugita et
al, 1998).  In fact, giant elliptical galaxies should be investigated
in  detail far beyond  $z=1$ (see  Cimatti, 2001,  these proceedings):
diagnostics from  the strong CaII  to MgI absorption systems  will set
constraints  on their  age  and  metallicity and  will  allow for  the
determination  of  their  stellar  content  and  the  epoch  of  their
formation.  These  lines are  redshifted in the  near-IR $I$,  $J$ and
$H$-bands. Detecting  the continuum  emission is essential  to measure
absorption lines.  The faintness  of the continuum emission in distant
objects  makes such  observations very  challenging.  They  require an
significant improvement  of the image quality  and limiting magnitude,
coupled with  3D spectroscopy.  Alternatively the first  epoch of star
formation can be dated in  massive galaxies hosting a quasar (Dietrich
and Hamann, 2001).  According to  Collin \& Joly (2000), heavy element
overabundances can not be  explained in the context of photoionisation
models and  are attributed to the  presence of a  starburst (Hamann \&
Ferland,  1993).  At  high redshifts  ($z=4.5$), quasars'  line ratios
(MgII/FeII) indicate  solar and  supersolar metallicities of  the gas,
which  could  be  related  to   SN  Ia  metal  production.   Given  an
evolutionary time scale  of $\sim$ 1Gy for the  progenitor stars of SN
Ia, this leads  to substantial star formation up  to $z=8$. Very first
events  of star  formation at  $z=10-20$  can be  probed from  deep
spectra of $z=6$ QSOs (see Fan  et al, 2001), for which the FeII lines
($<$3200\AA~  at  rest)  are  redshifted  in the  $J$  and  $H$-bands.
Improvements  of  image quality  and  moderately  high resolution  are
crucial for programs aimed  at probing bulges of distant galaxies, 
hence small, faint objects.
 
\subsection{Stellar physics: a clue to open questions in cosmology}

Type Ia  supernovae are the best extragalactic  standard candles known
so far.  The  Hubble diagram established from $z=0.5$  to $z=0.83$ has
already  shown that  cosmologies with  non-zero  cosmological constant
($\Lambda>$0)  are  favored by  the  observations  (Perlmutter et  al,
1998).   However, the value  of the  cosmological constant  has little
effect on  the Hubble diagram at  low redshift, given  the accuracy of
present-day observations.  This surprising result by Perlmutter et al.
therefore needs to be extended  at higher redshift, where the detailed
predictions of  $\Lambda$ cosmologies can  be compared with  the data.
 The  most  recent discovery  places the  highest
redshift supernova  at $z=1.7$  (Riess et al.   2001), but due  to the
faintness of the target, there is no spectroscopic confirmation of its
redshift.  Spectroscopic follow-up is essential in this field, whether
it be to provide redshift information or to derive physical parameters
of  the supernova.  At $z>1$,  most spectral  features relevant  to the
study of  supernovae are redshifted  in $J$ and  $H$-bands.  Obtaining
very deep spectra is at the  edge of the possibility for 8 meter class
telescopes.  Pushing further their  spectroscopic limits, even by only
1 magnitude  would place  the present and  future results on  a firmer
ground  and  settle  the  issue  of  the  value  of  the  cosmological
constant. Fainter point source detection is one of the primary goal of
the FALCON concept.\\

Red  supergiants are present  in populations  with ages  between about
10$^7$ and  about 10$^8$\,yrs.  Their predicted  colour and luminosity
distributions  are very  sensitive  functions of  age and  metallicity
during that time.   Obtaining proper colour-magnitude distributions of
red giants is a promising way for the determination of the recent
history of  starburst and young post-starburst  galaxies. 
AGB stars are dominant sources of near-IR light at post-starburst ages 
of 10$^8$  to about 2\,  10$^9$\,yrs. The distribution of carbon stars
among these traces the intermediate age star formation history and 
carbon star proportions indicate metallicity. Theoretical
spectra for luminous cool giants suggest that metallicity  has a very
strong effect on the spectra.  The systematic studies on near-IR metal
lines in globular cluster red  giants by Frogel et al. (1999) promises
that it  will become  possible to estimate  metallicities from  $H$ or
$K$-band  spectra (R=5000-10000)  even  from very  late type  spectra.
Once  these calibrations  are obtained  and  extended to  RSG and  AGB
stars, detailed  diagnostics will be available  for the post-starburst
galaxies and  massive intermediate age clusters that  FALCON shall observe
spectroscopically out  to $\sim\,100\,$Mpc. These  studies have direct
repercussions on distant galaxies  studies, because the calibration of
the stellar mass is far from being accurately established.

\section{Proposed specifications for a 2nd generation spectrograph at VLT}

{\bf  Going deep:  more object,  less sky.}  The S/N  performances are
limited by the  light concentration entering a spatial  element of the
spectrograph.  For star-like sources  the S/N is directly proportional
to $FWHM^{-2}$.  For  more extended faint sources the  key issue is to
minimize  the  ratio of  the  sky to  object  light  which enters  the
spectrograph. Since most of the  distant galaxies ($z> 0.5$) have half
light radius  $r_{05}$ in the  range 0".15 to  0".5, it is  crucial to
sample about one  tenth of an  arcsecond, i.e. 1-2 kpc, the size of a 
large star forming region. Adaptive optics coupled with spectroscopy is
 the ideal way to reach this goal.\\

Only 3 D spectroscopy  can address  the question  of galaxy
dynamics,   including    the   interacting   systems    and   galactic
disks. Spectroscopy  using integral field units with  a total aperture
adapted  to distant  galaxy  sizes will,  in  one shot  map the  whole
velocity  field:  a  goal  out   of  reach  of  the  traditional  slit
spectroscopy. \\
 
Going deep implies improving the overall throughput of the instrument.
It also implies that efforts should  be put on limiting the sky signal
to  the strict minimum.   OH suppression  techniques, whether  they be
under  the  form of  filtering  or  numerical  techniques, requires  a
spectral  resolution $R\ge  5000$.  Since  the velocity  dispersion of
most distant galaxies  spans over the range 20-100  km/s, the spectral
resolution  should  in  fact  exceed  this  and  reach  values  around
$R=15000$,  which  is  also  adequate  for studies  of  abundances  in
stars. \\

{\bf A wide field and multiplex capability.} A large field of view for
selecting targets in a multi-IFU  mode is a prerequisite.  This would
optimize the  efficiency of follow up observations  of surveys carried
out at other wavelengths over a large field of view, e.g., the sources
provided  by XMM,  SIRTF or  HERSCHEL.  The  number density  of high-z
galaxies  bright enough  to be  observed in  3D spectroscopic  mode is
relatively  small.   For  example  the  number  density  of  starburst
galaxies  detected  by  ISO at  $z\ge0.5$  is  less  than 50  per  100
$arcmin^{2}$ and only a fifth of  them would have enough flux in their
emission lines to allow for a 3D spectroscopic analysis
of  their velocity  fields.  Stellar  fields in  Local  group galaxies
require very large  fields of view.  Keeping a field  of view close to
that of the Nasmyth ($\Phi$= 25 arcmin) of the VLT units
is therefore capital.\\
  
{\bf High  spectral resolution: going for emission  lines.} A spectral
range extending to  the near-IR would allow to  take full advantage of
the adaptive optics technics and to observe always in sub-tenth-arcsec
seeing.   The  0.7-1.8$\mu$m domain  includes  most  of the  important
spectral  features for  galaxies  at $z=0.5-2$.   Optimal targets  for
dynamical studies of galaxies (3D  spectroscopy) are limited with an 8
meter telescope,  to $\sim$ $I_{AB}$=23.5  (or $H_{AB}$=22 for  an Scd
energy distribution) galaxies.  Most  of these galaxies have redshifts
in  the range  $z=0.5-2.5$, with  a peak  at $z\sim  1.5$.   Their gas
velocity field  can be mapped using  the $H\alpha$ line  up to $z=1.5$
and up to $z=2.5$ or more using the [OII]3727, $H\beta$ and [OIII]5007
lines.

\section{The FALCON concept and expected performances}

A major  limitation to traditional  adaptive optics is the  very small
field of view  inherent to the isoplanatic patch  size.  While this is
no serious  problem for detailed studies  of individual sources,  it is a
killer as soon as the aim is  to survey large number of objects, or to
follow up  sources detected  at other wavelengths  on a wide  field of
view.  It has been demonstrated  (Gendron et al, 2001, in preparation)
that, by relaxing several conditions of the adaptive optics, one could
significantly improve  image quality  in the near-IR,  while providing
corrections  for a  large  number of  objects  in wide  (cosmological)
fields. This can be done by:

\begin{itemize}

\item restricting the adaptive optics correction to 0".15-0".25 FWHM

\item  optimizing  the  correction  to  several  small  areas  of  few
$arcsec^{2}$ each,  and selected within  a large field (for  example a
Nasmyth field, $\Phi$= 25 arcmin).

\item using 3  reference stars (R$\le$ 16) per  small area (or target)
for a multi-analysis of the wavefront (MCAO technics).

\item considering low (3 to 5 Zernike modes) to moderately high ($\ge$
60 Zernike modes) order compensations.

\end{itemize}

\begin{figure}[t]
\begin{center}
\includegraphics[width=.8\textwidth]{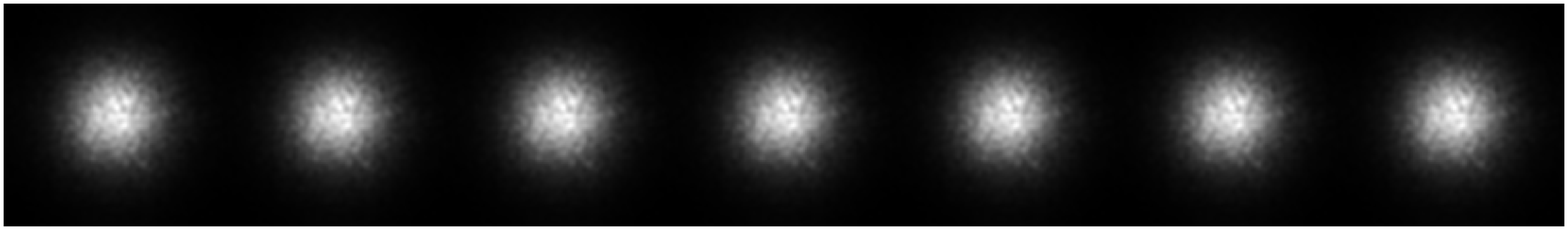}
\includegraphics[width=.8\textwidth]{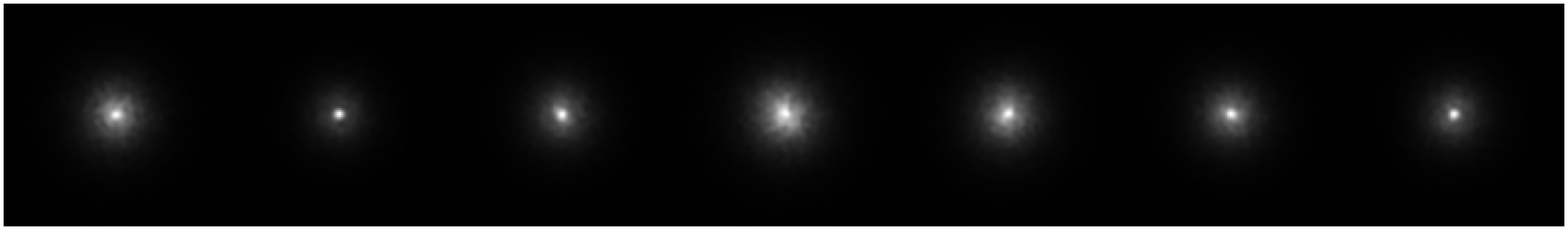}
\end{center}
\caption[]{Seven examples of  uncorrected stellar sources at locations
randomly selected in a cosmological field;({\it Bottom}): same sources
after  MCAO correction  with  optimization at  source location  (three
reference  stars) assuming a  correction up  to the  15$^{th}$ Zernike
order.}
\label{eps1}
\end{figure}

Simulations  have been  performed assuming  three turbulent  layers at
0km,  1km  and  10km, with  phase  variance  of  20\%, 60\%  and  20\%
respectively,  and  a resulting  seeing  of  0".65  at 5500\AA~.   500
galaxies   have  been   randomly  selected   within   five  $\Phi$=25'
cosmological fields ($b^{II}$$\ge$ 45 degrees), and the phase has been
estimated  from a MCAO  system simulator  developped by  Fusco (2000),
assuming a  Shack-Hartmann sensor type and  a S/N of  10 which roughly
corresponds to  R$\le$ 16 reference  stars. Figure 1 displays  how the
image  quality is improved  at different  object locations.   Figure 2
probes  that large  gains in  S/N  can be  obtained ($\sim$  1 mag  at
1.6$\mu$m)  by correcting only  the 5  first Zernike  modes (including
tip-tilt  and defocus).  Obtaining  a significant  further gain  would
requires higher order compensations up to 60 Zernike modes, but a gain
by 1 mag already allow to fulfill many proposed science drivers.\\

\begin{figure}[b]
\begin{center}
\includegraphics[width=.8\textwidth]{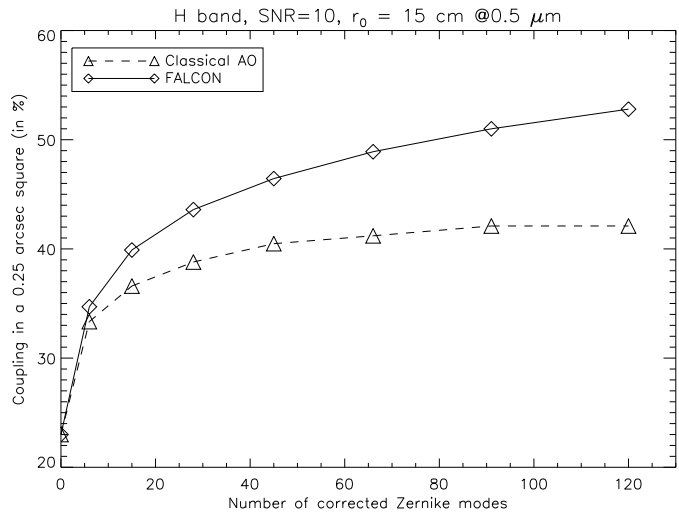}
\end{center}
\caption[]{Average fraction of light  entering a 0".25 square aperture
after MCAO correction as a function of the number of corrected Zernike
modes. Sources have been assumed  to be stellar and the correction was
done at 1.6$\mu$m }
\label{eps2}
\end{figure}

The basic  concept of  FALCON (Say\`ede et  al, 2001,  in preparation)
includes  small  miniaturized   devices,  called  "adaptive  buttons",
located   on  the   beam  entering   the  IFUs   across   the  Nasmyth
field. Similar  buttons will be  used on objects and  reference stars,
and allow for correction of  the wave-front. The development of FALCON
({\it Fiber-spectrograph with Adaptative-optics on Large-fields to Correct
 at Optical and Near-infrared}) will be done in two stages:

\begin{enumerate} 

\item tip-tilt  and defocus  with a  five
motorized  axis lens  (Figure 3)\\

\item  correction of  higher order
modes with a micro deformable mirror

\end{enumerate}

Only  a limited number  of reference  stars is  required in  the first
stage (1 star for  most of the targets, see Figure 2),  and for such a
limited correction,  the system could work  either in open  loop or in
partially  closed   loop.  Although  far  superior   to  any  existing
instrument by its  image quality, it would be  similar by many aspects
to  GIRAFFE  at  VLT.   GIRAFFE  will  provide  the  first  system  of
deployable multi-IFUs (20 IFUs, Figure  4) available at the focus of a
very large telescope.

\begin{figure}[t]
\begin{center}
\end{center}
\caption[]{JPEG Files , Example  of an  adaptive  button for  correcting 5  Zernike
modes (concept by F. Say\`ede  and F. Rigaud).  The lens-doublet L2 is
in the  pupil plane and L3 is  a field lens. Microlenses  in the image
plane of the button provide a spatial sampling of the source and adapt
the light injection in the otical fibers.}
\label{eps2}
\end{figure}

Several developments are underway for optimizing the correction system
(measurements   and  actuation),  and   the  control   loop.   Further
investigations  are required  to  adapt the  fiber  system to  FALCON
(including microlenses, $\sim$ 50$\mu$m diameter IR fibers, limitation
of  the  focal ratio  degradation  losses).   The second  stage  is more
  ambitious  and would  highly
benefit  of  the former  feasibility  study.  The  goal  is  to add  a
micro deformable mirror (MDM) to  the compact adaptive button  in order to correct
higher  order Zernike modes  (Figure 2), requiring a relatively high actuator
density. The main advantages of MDMs are their compactness, scalability,
and specific task customization using elementary building blocks, including
on-board electronics.   Four such  adaptive buttons
will be  needed for each target,  including 3 on  the reference stars,
for which a  close loop correction is needed.   The whole system would
eventually work in a partially  closed loop mode, under the control of
the MCAO software and taking advantage of the similarity between the 4
devices (study in progress).  The resulting instrument will eventually
produce   an   image   quality   competitive   with   that   of   NGST
spectrographs.\\

\begin{figure}[t]
\begin{center}
\end{center}
\caption[]{JPEG File, Face-on  view  of  one  of  the  20 GIRAFFE deployable IFUs
 (realisation by  L.  Jocou  and I. Guinouard, see also Jocou et al, 2000)
   with  20 microlenses
(size=140$\mu$m) adapted  on the prism;  performances ($\sim$ 60\%)
could be improved for FALCON by minimizing the focal ratio degradation
losses.}
\label{eps2}
\end{figure}

Assuming  the   completion  of  the   first  stage  (only   low  order
correction), FALCON would provide a S/N  gain of up to 1 magnitude for
a stellar  source, when compared  with an instrument  without adaptive
correction.  For  compact sources with  unresolved lines -such  as the
numerous HII regions in distant  galaxies, there will be an additional
gain following S/N $\sim$  EW/$\lambda$ x R, totalizing 2-3 magnitudes
improvement in  S/N if compared to R=2000  spectrographs, just because
the fraction of  spectra accessible ``between'' the sky  lines is much
larger.   For  example, at  R=10000,  FALCON  would  beat  NGST  at
scrutinizing the  properties of $\sigma$=  30km/s HII regions  at very
large distances.  A simple simulator  of FALCON, assumed to be mounted
at VLT, shows that for  a $L^{\star}$ galaxy at $z=1.5$ ($I_{AB}$=23.5
and  $H_{AB}$= 22  for  a Scd  spectral  type), the  S/N per  spectral
resolution element  (R=5000) at 1.6  $\mu$m would be 4.5$\pm$1  with a
total exposure time  of 2 hours. This number is boosted  to 130 for a
EW=60\AA~ $H\alpha$ line, and studies of the velocity field map could be
done in great details (Figure 5). FALCON would
be  unique for  spectroscopy of  supernovae at  $z=1.5$: if  we assume
$J_{AB}$=25  after   an  extrapolation   from  SN  1997ap   at  z=0.83
(Perlmutter et al, 1998), a  S/N=5$\pm$2 (R= 600) would be reach after
6 hour exposures.
   
\section{Summary}

\begin{figure}[t]
\begin{center}
\includegraphics[angle=-90,width=1\textwidth]{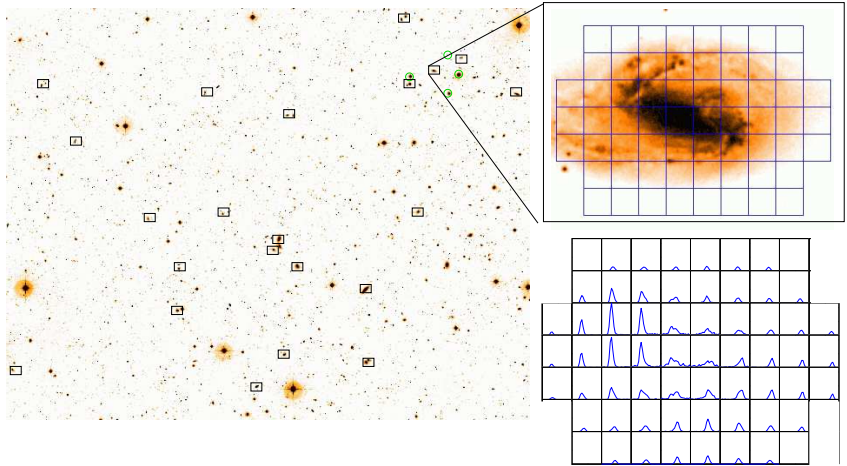}
\end{center}
\caption[]{Several  IFUs (squares)  in a  $\Phi$= 25  arcmin  field; 3
reference stars  are shown (circles)  for one of  them; a zoom  in the
upper  right  show  an  individual  IFU  made  with  62  0".12  square
microlenses  laid on a  $z=1.48$ $I_{AB}$=23.5  galaxy; on  the bottom
right  the   corresponding  spectra  around  the  EW=60\AA~ $H\alpha$
  line  is
presented, revealing the presence of an  HII region on the left of the
galaxy, as well as resolving its velocity field }
\label{eps3}
\end{figure}

We propose a  new concept for developing spectroscopic  studies with 8
meter ground based telescopes. FALCON would allow adaptive corrections
for many targets in large fields of view and would achieve significant
improvement of the image quality  or light concentration for more than
80\%  of the  cosmological sources.   A realistic  design would  be 60
deployable IFUs (Figure 5) within a $\Phi$= 25 arcmin field, providing
60  x  62 =  3720  spectra  covering (R=5000)  the  $I$,  $Z$, $J$  or
$H$-bands.  Spectral resolutions would range from R=5000 (required for
a  proper  removal  of  strong  OH  lines)  to  R=20000  (for  stellar
studies). The first stage of development for FALCON is to correct five
of the six Zernike modes (tip-tilt, defocus and 2 astigmatisms) with a
lens, providing  a 1 magnitude gain  in S/N for a  stellar source. The
second  stage  is aiming  at  correcting  higher  order modes  with  a
micro deformable mirror,  hence providing  a supplementary  gain of  1 
magnitude.
FALCON could be implemented at  the OzPoz at UT2 (Kueyen). Offering 3D
spectroscopy  and  medium resolving  power,  it  will  be the  natural
complement  of NGST  for dynamical  studies of  distant  galaxies, for
follow-up  studies of  XMM, SIRTF  and PLANCK  sources,  for abundance
surveys  of cold  stars  within and  beyond  the Local  Group and  for
massive  spectroscopic  studies  of stellar  populations  in
general.

%

\end{document}